\documentclass[aps, pre, reprint]{revtex4-1}
\usepackage{graphicx}
\usepackage{color}
\usepackage{amssymb}
\usepackage{amsmath}
\usepackage{tabularx}
\usepackage{bm}
\usepackage{hyperref}
\usepackage{ltxtable}
\usepackage{natbib} 
\usepackage{longtable}
\usepackage{float}
\newcommand{\be}{\begin{equation}}
\newcommand{\ee}{\end{equation}} 
\newcommand{\bea}{\begin{eqnarray}}
\newcommand{\eea}{\end{eqnarray}}

\usepackage[usenames,dvipsnames]{xcolor}
\hypersetup{
colorlinks,
citecolor=blue,
filecolor=blue,
linkcolor=blue,
urlcolor=blue}

\begin{document}
\title{Large eddy simulations of turbulent thermal convection using renormalized viscosity and thermal diffusivity}

\author{Sumit Vashishtha}
\email{vash.sumit@gmail.com}
\affiliation{Department of Physics, Indian Institute of Technology, Kanpur 208016, India}
\author{Mahendra K. Verma}
\email{mkv@iitk.ac.in}
\affiliation{Department of Physics, Indian Institute of Technology, Kanpur 208016, India}
\date{\today}

\begin{abstract}
In this paper we  employ renormalized viscosity and thermal diffusivity to construct a subgrid-scale model for  large eddy simulation (LES) of turbulent thermal convection. For  LES, we add $\nu_\mathrm{ren} \propto  \Pi_u^{1/3} (\pi/\Delta)^{-1/3}$ to the kinematic viscosity; here $\Pi_u$ is the turbulent kinetic energy flux, and   $\Delta$ is the grid spacing.  In our model, the turbulent Prandtl number is unity.  We performed LES of turbulent thermal convection on a $128^3$ grid and compare the results with direct numerical simulation (DNS) on a $512^3$ grid. There is a good agreement between the LES and DNS results on the evolution of kinetic energy and entropy, spectra and fluxes of velocity and temperature fields, and the isosurfaces of temperature.  We also show the capability of our LES to simulate thermal convection at very high Rayleigh numbers and exhibit some results for $\mathrm{Ra=10^{18}}$. 

\end{abstract}

\maketitle
\section{Introduction}
\label{sec:intro}

Turbulence is one of the most difficult phenomena to simulate on a computer due to vast range of length scales involved.    In a direct numerical simulation (DNS), all the length scales of the flow need to be resolved, which is very challenging for  large Reynolds numbers.  This problem is circumvented in large eddy simulations (LES) where the small-scale fluctuations are modelled. Thus, only the large and intermediate scales are resolved which makes LES computationally less expensive and practical compared to DNS.  

In hydrodynamic turbulence, the velocity field at  different scales interact with each other and create a cascade of energy, called energy flux $\Pi_u$.  The energy flux in the inertial regime equals the energy dissipation. Scaling analysis reveals that the effective viscosity at length scale $l$ is proportional to $\Pi_u^{1/3} l^{4/3}$; this viscosity enhances the diffusion of linear momentum.  This feature is exploited in eddy-viscosity based subgrid-scale (SGS) models of LES.

The earliest SGS model was proposed by Smagorinsky~\cite{smagorinsky1963general} who modelled the effective viscosity as,
 \be
 \nu_\mathrm{Smag} = (C_s \Delta)^2 \sqrt{2 \bar{S}_{ij} \bar{S}_{ij}},
 \label{eq:Smag}
 \ee
 where $\bar{S}_{ij}$ is the stress tensor at the resolved scales, $\Delta$ is the smallest grid scale, and $C_s$ is a constant that is taken between 0.1 and 0.2.   A less popular but theoretically  rigorous LES model is based on the renormalised viscosity.  Using  renormalisation group (RG) analysis,~\citet{Yakhot:JSC1986,McComb:PRA1992Two_field,McComb:book:Turbulence,McComb:book:HIT, Zhou:PRA1988,Zhou:PR2010} estimated the effective viscosity.  In one of the computations, \citet{McComb:PRA1992Two_field,McComb:book:Turbulence,McComb:book:HIT} showed that the renormalized viscosity is
  \be
\nu_\mathrm{ren}(k) = K_\mathrm{Ko}^{1/2} \Pi^{1/3} k^{-4/3}\nu_*,
\label{eq:nu_r0}
\ee
where $K_\mathrm{Ko}$  is the Kolmogorov's constant, and $\nu_*$ is a constant. 
Using RG computation,~\citet{McComb:PRA1992Two_field} found that $\nu_* \approx 0.50$ and $K_\mathrm{Ko} \approx 1.62$,  while ~\citet{Verma:PR2004}  computed the above quantities using a refined technique and found $\nu_* \approx 0.38$ and $K_\mathrm{Ko} \approx 1.6$. In this approach, it is assumed that the length scale corresponding to the grid spacing $\Delta$ lies in the inertial range where the energy spectrum $E_u(k) \sim k^{-5/3}$,  and the effective viscosity follows Eq.~(\ref{eq:nu_r0}) with $k=k_c = \pi/\Delta$. Refer to  \citet{Verma:Pramana2004} and \citet{Vashishtha:LES_fluid_arXiv}  for LES of hydrodynamic turbulence using renormalized viscosity.

Turbulent thermal convection is more complex than hydrodynamic turbulence due to the presence of another field (temperature) and thermal plates. Owing to the  uncertainty in the model of turbulent convection, an effective SGS model for such flows have eluded engineers and scientists. However, there have been a few attempts in this direction. Eidson~\cite{eidson1985numerical} extended the Smagorinsky's eddy-viscosity based model  to turbulent thermal convection. \citet{Jie:LES_2D_RBC} used a similar approach for LES of 2D Rayleigh-B\'enard convection. To overcome the excessive dissipation added in Smagorinsky's eddy viscosity model, Wong and Lilly~\cite{Wong:PF1994} employed dynamic LES to turbulent convection. However, dynamic LES  itself has   numerical instabilities issues due to  spatial averaging during the  evaluation of model parameters. Foroozani et al.~\cite{Foroozani:PRE2017} overcome these issues by employing Lagrangian dynamic subgrid-scale model~\cite{meneveau_lund_cabot_1996}  and studied reorientations of the large scale structures in turbulent thermal convection. Besides these, a non-eddy viscosity model was constructed by Kimmel and Domaradzki~\cite{kimmel2000large} wherein the SGS quantities are estimated by expanding the temperature and velocities to scales smaller than the grid size. 

In this paper we construct an SGS model for LES of turbulent thermal convection using renormalized parameters.  Researchers~\cite{Yakhot:JSC1986} have performed RG computation of passive scalar turbulence, but its applicability to turbulent thermal convection is  highly debatable due to additional complexities~\cite{Lohse:ARFM2010, Verma:NJP2017}.  In thermal convection, buoyancy drives the flow, and the mean temperature gradient affects the thermal fluctuations in a nontrivial manner.  Also, thermal convection is anisotropic due to the buoyancy direction, in contrast to the hydrodynamic turbulence which is statistically isotropic in the inertial regime~\cite{Lohse:ARFM2010, Verma:NJP2017}. 

\citet{Lvov:PRL1991}, \citet{Lvov:PD1992}, and \citet{Rubinstein:NASA1994} employed field-theoretic tools to model turbulent thermal convection and argued that its kinetic energy spectrum follows Bolgiano-Obukhov scaling, i.e., $E_u(k) \sim k^{-11/5}$, and $\nu(k) \sim k^{-8/5}$. Recent theoretical arguments and numerical simulations~\cite{Kumar:PRE2014, Verma:NJP2017}, however, show that turbulent thermal convection has properties similar to the hydrodynamic turbulence, i.e., $E_u(k) \sim k^{-5/3}$, and $\nu(k) \sim k^{-4/3}$.  \citet{Nath:PRF2016} and \citet{Verma:NJP2017} also show that turbulent thermal convection is nearly isotropic, and the energy transfers in such flows are local and forward. We construct a LES of turbulent thermal convection based on these observations.

Due to the aforementioned similarities between the hydrodynamic turbulence and turbulent thermal convection, we  employ renormalized viscosity of the form of Eq.~(\ref{eq:nu_r0}) to turbulent convection as well.  Though temperature field in thermal convection has relatively complex behaviour, yet, for simplicity, we take $\kappa(k) = \nu(k)$, or turbulent Prandtl number $\mathrm{Pr}(k) =1$. 

We perform DNS on a $512^3$ grid and LES on $128^3$ grid with the aforementioned renormalized parameters. This paper contains a detailed comparison between the DNS and LES results.  We show that  the evolution of the total kinetic energy and entropy, as well as the spectra and fluxes of the temperature and velocity fields of DNS and LES are approximately same. The large scale features of thermal plumes are captured quite well by our LES. Furthermore, we were able to simulate thermal convection at Rayleigh numbers (Ra) as high as $10^{18}$ and beyond using the present approach. These results indicate that our LES model is quite good for simulating turbulent thermal convection.
 
The outline of the paper is as follows: In Sec.~\ref{sec:Maths}  we detail our SGS model for the LES of turbulent thermal convection.  Simulation details are discussed in Sec.~\ref{sec:CM}. Results obtained from the LES and DNS are compared in Sec.~\ref{sec:Results}. In  Sec~\ref{sec:very_high_Ra} we describe some of the LES results for very high $\mathrm{Ra}$ numbers. We summarize our results in Sec.~\ref{sec:Conclusion}.
     
\section{LES formulations using renormalized parameters}
\label{sec:Maths}
 We consider a Boussinesq fluid kept between two horizontal plates that are separated by a distance $d$. The temperature difference between the two plates is $\Delta T $. This system, called Rayleigh-B\'{e}nard convection (RBC), is described by the following equations~\cite{Chandrasekhar:book:Instability}:
\bea
\frac{\partial{\bf u}}{\partial t}+({\bf u\cdot\nabla){\bf u}} & = & -\frac{1}{\rho_0}\nabla\sigma+\alpha \textsl{g}\theta\hat{z}+ \nu\nabla{}^{2}{\bf u},\label{eq:u_dim}\\
\frac{\partial\theta}{\partial t}+({\bf u\cdot\nabla)\theta} & = &\frac{(\Delta T)}{d} u{}_{z}+ \kappa\nabla{}^{2}\theta,\label{eq:th_dim}\\
\nabla\cdot{\bf u} & = & 0, \label{eq:div_dim}
\eea
where $\bf{u}$ is  the  velocity  field, $\theta$ and $\sigma$ are the  temperature  and  pressure  fluctuations  from  the conduction  state  respectively, and $\hat{z}$ is  the  buoyancy  direction.  Here $\alpha$ is  the  thermal  expansion coefficient, \textsl{g} is the acceleration due to gravity, and
$\rho_0$, $\nu$, $\kappa$ are the mean density, kinematic viscosity, and thermal diffusivity of the fluid respectively. 
We nondimensionalize Eqs.~(\ref{eq:u_dim}, \ref{eq:th_dim}, \ref{eq:div_dim}) using the temperature difference between the two plates $(\Delta T)$ as the temperature scale, the plates separation $d$ as the length scale, and $\sqrt{\alpha \textsl{g}{(\Delta T) d}}$ as the velocity scale. This yields the following system of equations:
\bea
\frac{\partial{\bf u}}{\partial t}+({\bf u\cdot\nabla){\bf u}} & = &-\nabla\sigma+\theta\hat{z}+ \sqrt{\frac{\mathrm{Pr}}{\mathrm{Ra}}}\nabla{}^{2}{\bf u},\label{eq:u_ndim}\\
\frac{\partial\theta}{\partial t}+({\bf u\cdot\nabla)\theta} & = & u{}_{z}+\frac{1}{\sqrt{\mathrm{Ra}\mathrm{Pr}}}\nabla{}^{2}\theta,\label{eq:th_ndim}\\
\nabla\cdot{\bf u} & = & 0,\label{eq:inc_ndim}
\eea
where the two non-dimensional parameters are the Prandtl number
\be
\mathrm{Pr}=\frac{\nu}{\kappa},\label{eq:pr}
\ee
and the Rayleigh number
\be
\mathrm{Ra} =\frac{\alpha {g}(\Delta T) d^{3}}{\nu\kappa}.\label{eq:Ra}
\ee

Representation of flow properties at various scales is more convenient in Fourier space. Using the definition of the Fourier transform,
\bea
\label{eq:u_Fr}
\mathbf{u}(\mathbf{x})&=& \sum_{\mathbf{k}}\mathbf{\hat{u}}(\mathbf{k})e^{i\mathbf{k.x}}, \\
\label{eq:theta_Fr}
\theta(\mathbf{x})&=& \sum_{\mathbf{k}}\mathbf{\hat{\theta}}(\mathbf{k})e^{i\mathbf{k.x}},
\eea
we derive the RBC equations in Fourier space as:
\small
\bea
\label{eq:Fourier_mom}
\frac{d}{dt}\mathbf{\hat{u}(k)} + \mathbf{\hat{N}_u(k)} & = & -i\frac{1}{\rho_0}\mathbf{k}\hat{\sigma}(\mathbf{k}) +\alpha\textsl{g}\theta(\mathbf{k})\hat{z} - k^2\nu\mathbf{\hat{u}(k)} \\
\label{eq:Fourier_thermal}
\frac{d}{dt}{\hat{\theta}}(\mathbf{k}) + {\hat{N}_{\theta}(\mathbf{k})} & = & \frac{(\Delta T)}{d}\hat{u}_z(\mathbf{k})  - k^2\kappa\hat{\theta}\mathbf{(k)} \\
\label{eq:divergence_Fourier}
\mathbf{k}.\mathbf{\hat{u}(k)} & = & 0,
\eea
\normalsize
where the nonlinear terms are
\bea
\label{eq:N_u}
\mathbf{\hat{N}_u(k)}&=&\sum_{\mathbf{p}}\big[\mathbf{k} \cdot \mathbf{\hat{u}(q)}\big]\mathbf{\hat{u}(p)}, \\
\label{eq:N_theta}
\hat{N}_{\theta}(\mathbf{k})&=&\sum_{\mathbf{p}}\big[\mathbf{k} \cdot \mathbf{\hat{u}(q)}\big] \hat{\theta}(\mathbf{p}),
\eea
with $\mathbf{p+q= k}$.   The nonlinear terms of Eqs.~(\ref{eq:N_u}, \ref{eq:N_theta}) represent the triadic interactions among the wavenumbers $\mathbf{(k,p,q)}$ that satisfies $\mathbf{p+q= k}$, and are numerically computed using fast Fourier transforms (FFT).  In Fourier space, the pressure is computed using
\be
\hat{\sigma}(\mathbf{k}) = \frac{i}{k^2 } [{\bf k \cdot \hat{N}}_u({\bf k}) - \alpha g k_z \hat{\theta}({\bf k}) ].
\ee
In renormalisation group (RG) analysis of fluid turbulence, the Fourier modes of wavenumber shells are truncated iteratively~\cite{Yakhot:JSC1986,McComb:book:Turbulence,McComb:PRA1992Two_field,McComb:book:HIT, Zhou:PRA1988,Zhou:PR2010}  that leads to the elimination of some of the triadic interactions.  In RG procedure,  these eliminated interactions are taken into account by an enhanced viscosity. For hydrodynamic turbulence, It has been shown that the total effective viscosity at wavenumber $k$ is,
\be
\nu(k)  = \nu +  \nu_\mathrm{ren}(k) =\nu + K_\mathrm{Ko}^{1/2} \Pi_u^{1/3}k_c^{-4/3}\nu_*,
\label{eq:nu_total_RG}
\ee
where $\nu_\mathrm{ren}(k)$ is the renormalised viscosity that is added to the original kinematic viscosity.   The above derivation assumes Kolmogorov's spectrum for energy:
\be
E_u(k) = K_\mathrm{Ko} \Pi_u^{2/3} k^{_5/3}.
\ee
The equation for the energy flux  yields the Kolmogorov's constant as approximately 1.6.  

For passive scalar, \citet{Yakhot:JSC1986, Verma:IJMPB2001} performed renormalization group analysis and deduced that
\bea
\kappa(k)  & = & \kappa +  \kappa_\mathrm{ren}(k) =\kappa + K_\mathrm{Ko}^{1/2} \Pi_u^{1/3}k_c^{-4/3}\kappa_*,
\label{eq:kappa_total_RG} \\
E_\theta(k) & = & \mathrm{Ba} \Pi_\theta \Pi_u^{-1/3} k^{-5/3},
\eea
where $\kappa_* \approx 0.85$, and the Batchelor's constant $ \mathrm{Ba} \sim 1$. Thermal convection, however, is more complex than the turbulence dynamics of a passive scalar.  \citet{Kumar:PRE2014, Verma:NJP2017} showed that the kinetic energy spectrum of turbulent thermal convection is very similar to that of hydrodynamic turbulence ($\sim k^{-5/3}$), but the temperature field exhibits bispectrum with one branch as $k^{-2}$~\cite{Verma:NJP2017,Pandey:PF2016}.  \citet{Verma:NJP2017} and \citet{Nath:PRF2016}  showed that turbulent thermal convection is  isotropic in Fourier space, and that the energy transfers in Fourier space is local and forward, similar to that in hydrodynamic turbulence. \citet{Borue:JSC1997} arrived at similar conclusions in their  analysis. Though there have been several attempts on field-theoretic treatment of thermal convection~\cite{Lvov:PRL1991, Lvov:PD1992,Rubinstein:NASA1994}, there is no rigorous renormalisation group analysis of turbulent thermal convection that is consistent with the observations of \citet{Kumar:PRE2014, Verma:NJP2017}.

Motivated by the numerical observations of \citet{Kumar:PRE2014} and \citet{Verma:NJP2017} that the properties of turbulent thermal convection are very similar to hydrodynamic turbulence, we model the viscosity in turbulent thermal convection as in Eq.~(\ref{eq:nu_total_RG}).   The spectrum of the  temperature field is quite complex, yet, for simplicity we assume that 
\be
\kappa(k) = \nu(k)
\ee
or that the turbulent Prandtl number is unity.  Using numerical simulations, we show that the above model works very well for turbulent thermal convection. 

For our LES scheme, we employ sharp spectral filter at cutoff wavenumber ${k_c}$:
\bea
\mathbf{\hat{\bar{u}}}(\mathbf{k}) &=& H(k_c - k)\mathbf{\hat u}(\mathbf{k}), \\
\mathbf{\hat{\bar{\theta}}}(\mathbf{k}) &=& H(k_c -k)\mathbf{\hat \theta}(\mathbf{k}), 
\eea
where ${H}$ represents Heaviside function, and ${k}_c = \pi/\Delta$, where $\Delta$ is the subgrid cutoff in real space. Under this scheme, the real space velocity and temperature fluctuations are
\bea
\label{eq:H_u}
\mathbf{\bar{u}}(\mathbf{x}) &=& \sum_{\mathbf{k}}e^{i\mathbf{k.x}}\mathbf{\hat{\bar{u}}}(\mathbf{k}) = \sum_{|\mathbf{k}|<|\mathbf{k_c}|}e^{i\mathbf{k.x}} \mathbf{\hat{{u}}}(\mathbf{k}),  \\
\label{eq:H_theta}
\mathbf{\bar{\theta}}(\mathbf{x},t) &=& \sum_{\mathbf{k}}e^{i\mathbf{k.x}}\mathbf{\hat{\bar{\theta}}}(\mathbf{k}) =\sum_{|\mathbf{k}|<|\mathbf{k_c}|}e^{i\mathbf{k.x}}\mathbf{\hat{{\theta}}}(\mathbf{k}).
\eea
Under these assumptions, the equations for the resolved Fourier modes are:
\small
\bea
\label{eq:Fourier_mom_reduced}
\frac{d}{dt} \mathbf{\hat{\bar{u}}(k)} + \mathbf{\hat{\bar{N'}}_u(k)} & = & -i\mathbf{k}\frac{1}{\rho_0} \hat{\bar{\sigma}}(\mathbf{k}) +\mathbf{\hat{\bar{\theta}}}(\mathbf{k}) \hat{z} - k^2 \nu_\mathrm{tot} \mathbf{\hat{\bar{u}}(k)}, \\
\label{eq:Fourier_thermal}
\frac{d}{dt}\hat{\bar{\theta}}(\mathbf{k}) + \hat{\bar{N'}}_{\theta}(\mathbf{k}) & = & \frac{(\Delta T)}{d}\hat{\bar{u}}_z(\mathbf{k})  - k^2\kappa_\mathrm{tot} \hat{\bar{\theta}}\mathbf{(k)}, \\
\label{eq:div_Fourier}
\mathbf{k}.\mathbf{\hat{\bar{u}}(k)} & = & 0,
\eea
\normalsize
where,
\bea
\label{eq:N_u_Fr}
 \mathbf{\hat{\bar{N'}}_u(k)} &=&\sum_{\mathbf{|k|,|p|,|q|}<k_c}\big[\mathbf{k} \cdot  \mathbf{\hat{\bar{u}}(q)}  \big] \mathbf{\hat{\bar{u}}(p)},  \\
\label{eq:N_theta_Fr}
 \hat{\bar{N'}}_{\theta}(\mathbf{k})  &=&\sum_{\mathbf{|k|,|p|,|q|}<k_c}\big[\mathbf{k'} \cdot  \mathbf{\hat{\bar{u}}(q)}  \big] \hat{\bar{\theta}}(\mathbf{k''})
\eea
with ${\bf k =  p+q}$. As discussed above, for LES, we take
 \bea
\nu_\mathrm{tot}  & = & \nu +  \nu_\mathrm{ren}(k_c) =\nu + K_\mathrm{Ko}^{1/2} \Pi_u^{1/3}k_c^{-4/3}\nu_*,
\label{eq:nu_total_LES}  \\
\kappa_\mathrm{tot}  & = & \nu_\mathrm{tot}, \label{eq:kappa_total_LES} 
\eea
where $k_c = \pi/\Delta$ with $\Delta$ as the grid spacing, which is uniform in our simulation.

Now, several important issues regarding LES implementation are in order.  The computation of $\nu_\mathrm{tot}$ for LES requires the kinetic energy flux $\Pi_{u}(k_0)$, where $k_0$ is in the inertial range.  In our simulations, we compute $\Pi_{u}(k_0)$ using the  formula proposed by \citet{Verma:PR2004} and ~\citet{Dar:PD2001}:
 \be
 {\Pi_u{(k_0)} = \sum_{k\geq k_0}\sum_{p< k_0}\delta_{k,p+q}\mathrm{Im}[\mathbf{k.u(q)}][\mathbf{u^{*}(k).u(p)}]}.
 \label{eq:Pi}
\ee

Regarding the choice of $k_c$ in an $N^3$ box simulation, we take $k_c=2\pi N/3$ due to dealaising employed in our DNS and LES. Under the 2/3 rule of dealaising, the Fourier modes $|{\bf k}| > 2\pi N/3$ are set to zero. Hence, the nonzero Fourier modes are $k_i = [-2\pi N/3:2\pi N/3]$, where $i=x,y,z$.  Therefore, the effective $k_\mathrm{max} = 2\pi N/3$. We employ the above $k_c$ in our LES.

In Sec~\ref{sec:CM} we discuss the details of our numerical simulations.

\section{Simulation details}
\label{sec:CM}

We employ pseudo-spectral method for our numerical simulations and solve Eqs.~(\ref{eq:Fourier_mom}-\ref{eq:divergence_Fourier}) for DNS, and Eqs.~(\ref{eq:Fourier_mom_reduced}-\ref{eq:div_Fourier}) for LES.  We use the convection module of the code Tarang~\cite{Verma:Pramana2013tarang, Chatterjee:JPDC2018} to perform DNS on a $512^3$ grid, and LES on a $128^3$ grid.  For our simulations, we employ free-slip and conducting boundary conditions at the top and bottom walls, and periodic boundary condition at the side walls.  The box size is taken to be unity.  We  time advance  the equations using fourth-order Runge-Kutta method, and employ $2/3$ rule~\cite{Canuto:book:SpectralFluid} for dealiasing; the Courant-Friedrichs-Lewy (CFL) condition is used to determine the time step ${\Delta t}$.  We perform our simulations till a steady state is reached.

\begin{figure}
\begin{center}
\includegraphics[scale=0.35]{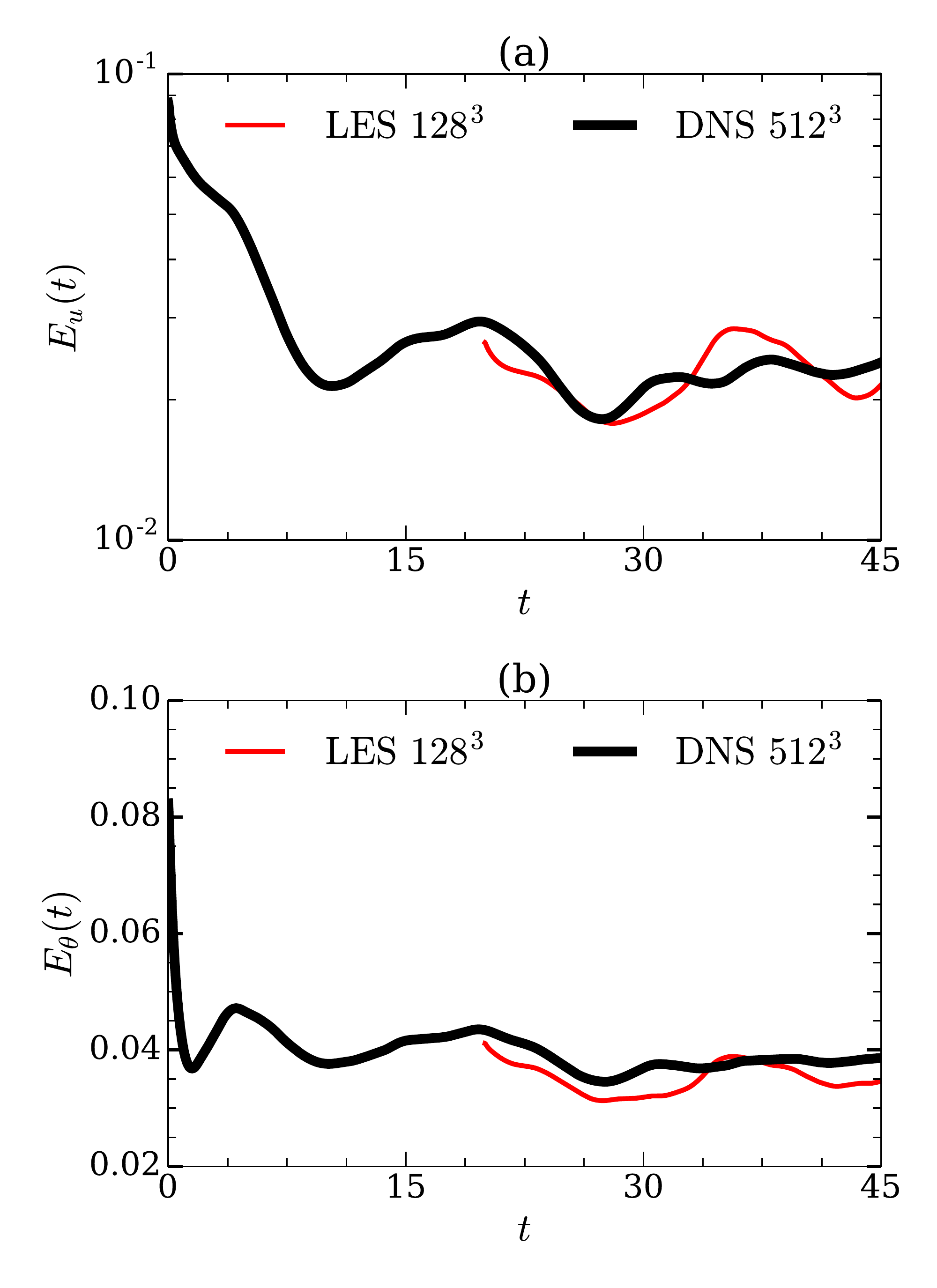}
\end{center}
\caption{For LES on a $128^3$ grid and DNS on a $512^3$ grid for $\mathrm{Ra} = 10^8$, temporal evolution of (a) total kinetic energy $E_u(t)$, and (b) total entropy $E_{\theta}(t)$. The LES was started using the truncated data of DNS at $t=20$.  The LES and DNS results match quite well.}
\label{fig:KE_PE} 
\end{figure}
 
 We perform our DNS and LES for $\mathrm{Ra=10^8}$.  We evolve our DNS from $t=0$ to $t=20$.  At this point, LES on a $128^3$ grid is turned on (see Fig.~\ref{fig:KE_PE}). We use the steady-state flow profile of DNS at $t=20$ as an initial condition for LES;  here we employ a spectral reduction of  $512^3$ grid data to a $128^3$ grid. Thus, the initial Fourier modes of the LES (at the resolved scales) are exactly same as those in DNS at $t=20$.  The original  DNS on $512^3$ is continued beyond $t=20$ along with the LES.


\begin{figure}
\begin{center}
\includegraphics[scale=0.35]{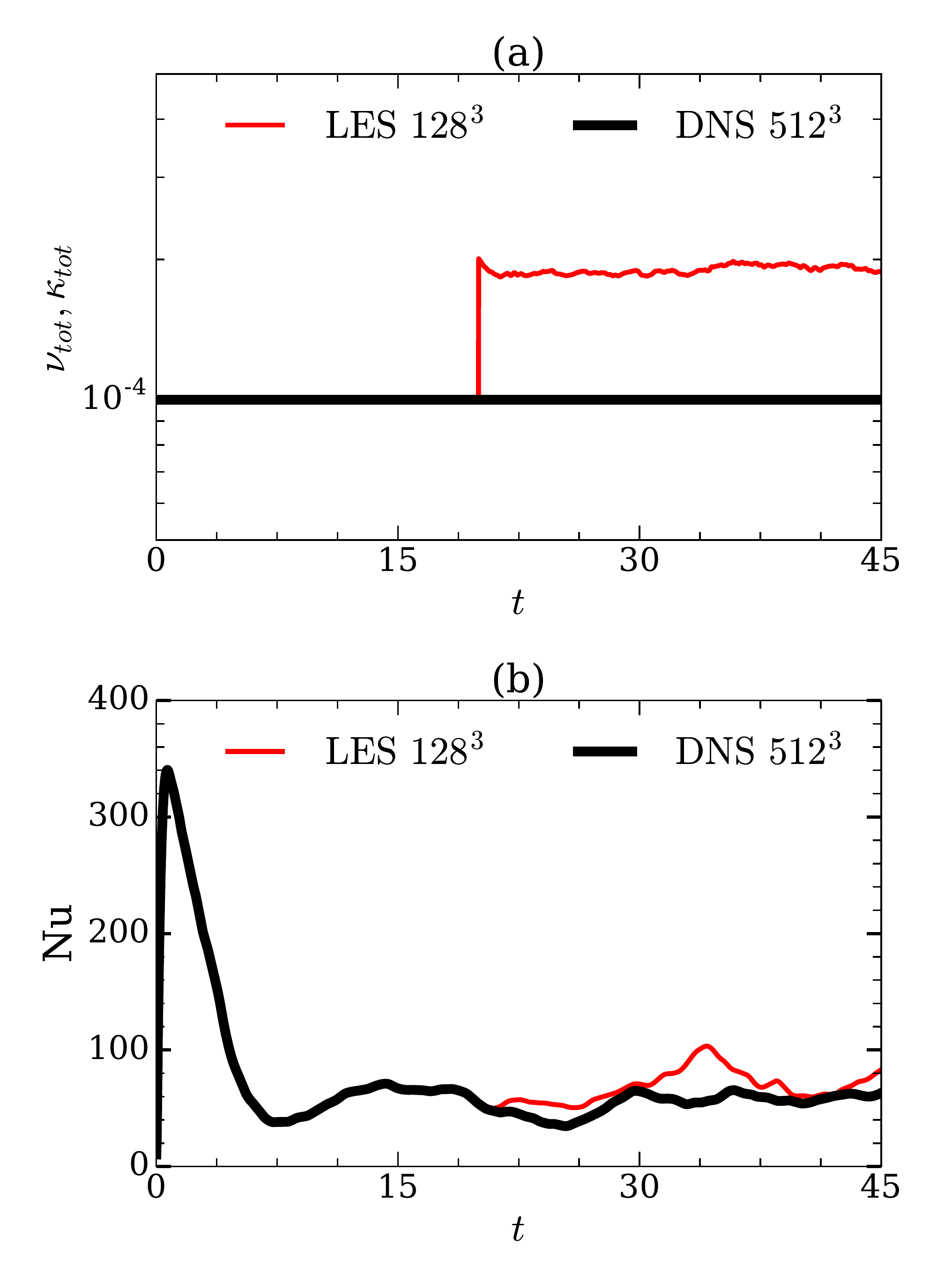}
\end{center}
\caption{For LES and DNS, temporal evolution of (a) total viscosity $\nu_\mathrm{tot}$, and (b) Nusselt number $\mathrm{Nu}$.  Note that $\nu_\mathrm{tot} = \nu$ for DNS.}
\label{fig:Nu_nu} 
\end{figure}
 
Note that for LES, we take $k_c=2\pi N/3$ with $N=128$,  and the viscosity and thermal diffusivity as  in Eq.~(\ref{eq:nu_total_LES}, \ref{eq:kappa_total_LES}).  For  DNS, 
\be
\nu_\mathrm{tot} = \kappa_\mathrm{tot}  = \nu = \kappa = \sqrt{\mathrm{Pr/Ra}} = 10^{-4}.
\ee  The simulations are continued till 45 non-dimensional time units; here the time unit is $L/U$, where $L,U$ are the large  length  and velocity scales respectively.
 
In the following section we compare the results of DNS and LES.

\section{Comparison of DNS and LES Results}
\label{sec:Results}

In this section, we compare the DNS and LES results on the evolution of global quantities such as total kinetic energy ($u^2/2$) and entropy ($\theta^2/2$).  We also compare the spectra and fluxes of the kinetic energy and entropy, as well as the isosurfaces of temperature. 

\begin{figure}
\begin{center}
\includegraphics[scale=0.35]{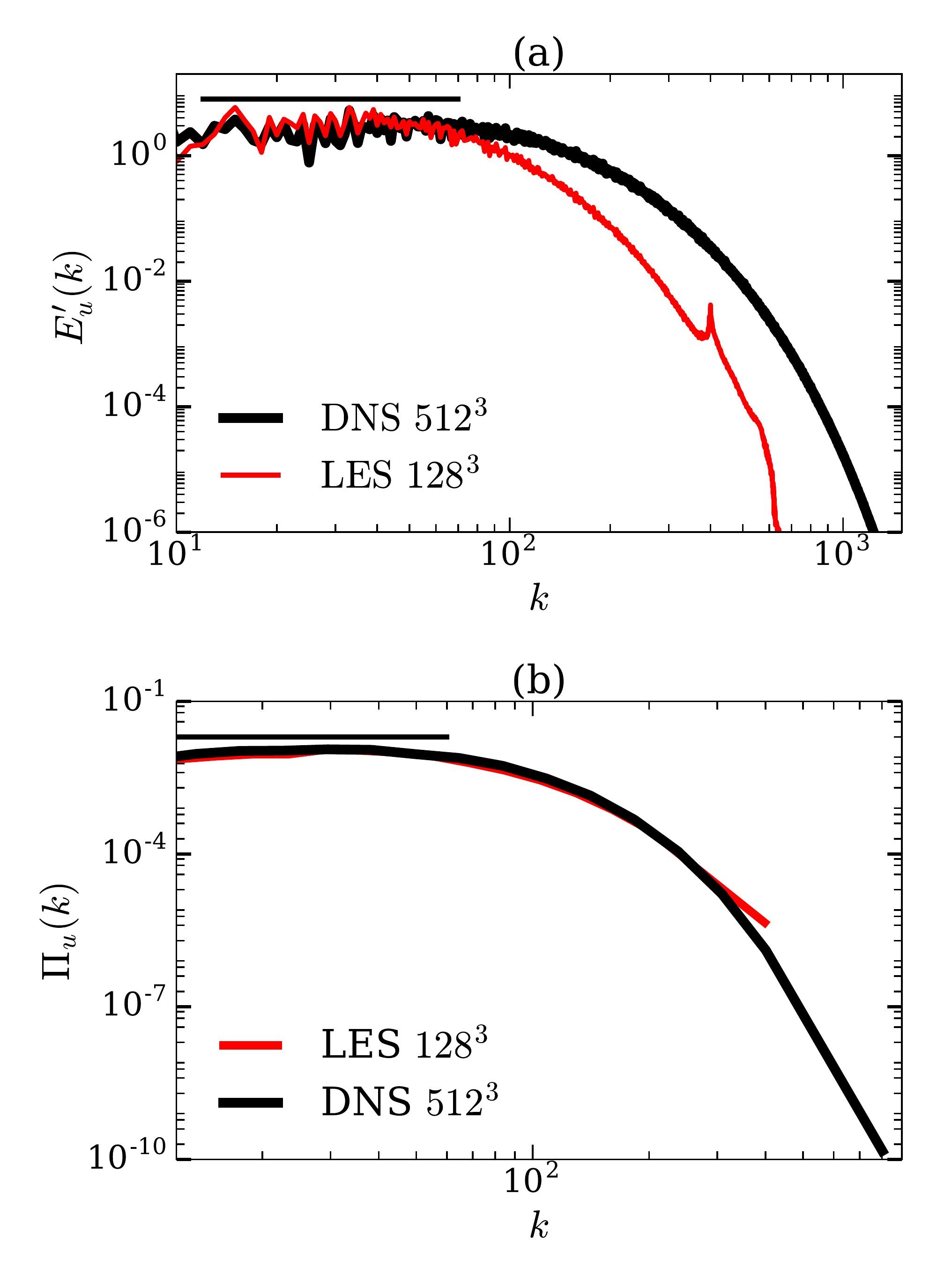}
\end{center}
\caption{For DNS and LES of RBC with $\mathrm{Ra}=10^8$, (a) Normalized Kinetic Energy spectrum ${E'_u(k)=E_u(k)k^{5/3}\Pi^{-2/3}}$, (b) Kinetic energy flux $\Pi_u(k)$. We observe these quantities to be approximate constants (see flat line) in the inertial range $k=[10,70]$. }
\label{fig:E_s_pi_1e8} 
\end{figure}

We start with the evolution of total energy ($E_u(t)$) and entropy ($E_{\theta}(t)$) which are defined as,
\small
\bea
\label{eq:E_u(t)}
 E_{u,\mathrm{DNS}}(t) &=& \frac{1}{2}\sum_{k}{|{\bf u}(\mathbf{k})|}^2;~E_{u,\mathrm{LES}}(t) = \frac{1}{2}\sum_{k}|{\hat{\bar{\bf u}}}(\mathbf{k})|^2, \\
 \label{eq:entropy}
 E_{\theta,\mathrm{DNS}}(t) &=& \frac{1}{2}\sum_{k}{|{\theta}(\mathbf{k})|}^2;~ E_{\theta,\mathrm{LES}}(t) = \frac{1}{2}\sum_{k}{|\hat{\bar{\theta}}(\mathbf{k})|}^2.
\eea
\normalsize
 In Fig.~\ref{fig:KE_PE}(a,b), we exhibit ${E_u(t)}$ and $E_{\theta}(t)$  for DNS and LES. We observe that $E_u(t)$ and $E_{\theta}(t)$ for DNS and LES are very similar (for $t>20$). Note however that the initial energy for LES  is slightly smaller than DNS.  This is because of the lesser number of modes in LES. In Fig.~\ref{fig:Nu_nu}(a,b)  we show the time series for $\nu_{\mathrm{tot}}$ and Nusselt number $\mathrm{Nu}$, which is the ratio of total heat flux (convective and conductive) to conductive heat flux:
\be
\mathrm{Nu}=\frac{\kappa{(\Delta T)}/{d} + {\big\langle u_z\theta \big\rangle}_{V}}{\kappa{(\Delta T)}/{d}}=1+{\big\langle u'_{z}\theta'\big\rangle}_{V},
\ee
where ${\big\langle \big\rangle}_V$ represents volume average, $u'_{z}=u_{z}d/{\kappa}$ and $\theta'=\theta/{(\Delta T)}$. 
Note that $\nu_{\mathrm{tot}}$ for LES is larger than $\nu$ of DNS. This is because of the additional  viscosity added to $\nu$ for LES (see Eq.~(\ref{eq:nu_total_LES})). 

\begin{figure}
\begin{center}
\includegraphics[scale=0.35]{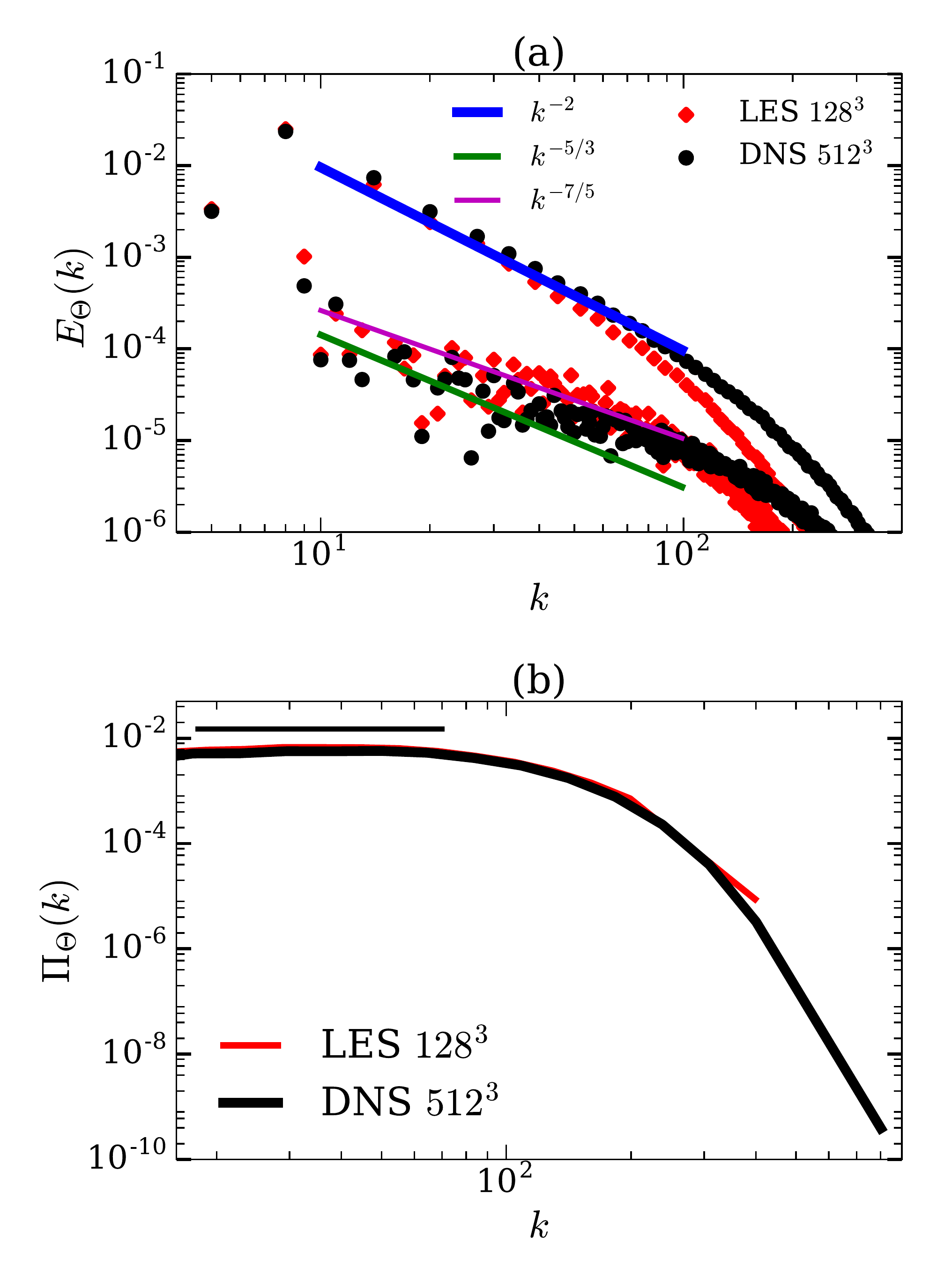}
\end{center}
\caption{For LES and DNS of RBC with $\mathrm{Ra}=10^8$, (a) Entropy spectrum $E_{\theta}(k)$ exhibiting bi-spectra and (b) entropy flux $\Pi_{\theta}(k)$. The flux is constant in the inertial range.}
\label{fig:E_th(k)_pi_1e8} 
\end{figure}

\begin{figure*}
\begin{center}
\includegraphics[scale=0.45]{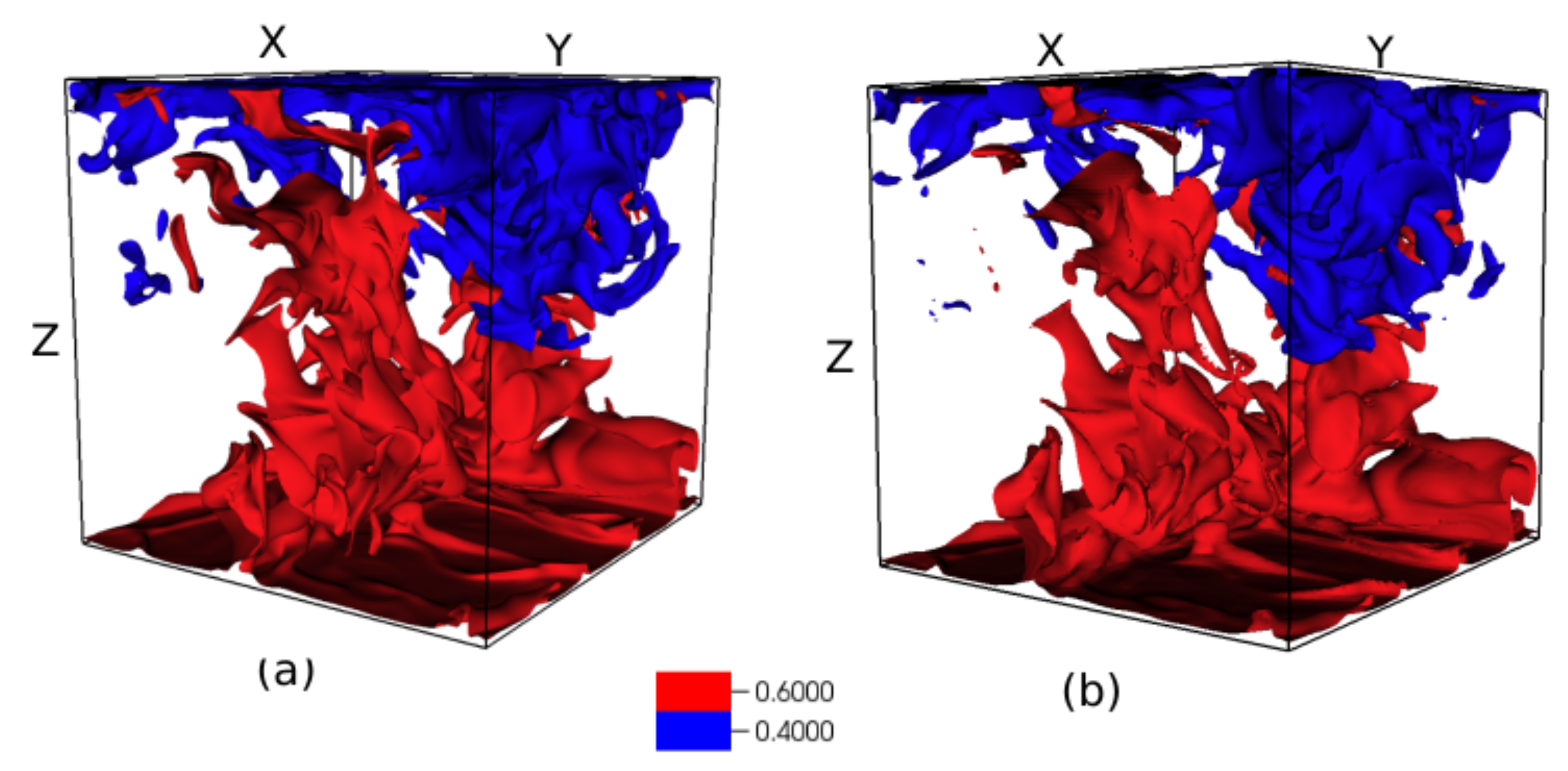}
\end{center}
\caption{For LES and DNS of RBC with $\mathrm{Ra}=10^8$: At $t=27$, the temperature isosurfaces obtained using the data of  (a) DNS and (b) LES. Note the similarity between large scale structures in the two plots. }
\label{fig:iso_1e8} 
\end{figure*}

Fig.~\ref{fig:E_s_pi_1e8}(a,b) exhibits the normalized kinetic energy spectrum $E'_u(k) = E_u(k)k^{5/3}\Pi_u^{2/3}$ and the kinetic energy flux ${\Pi_u(k)}$ at $t=30$, which is 10 time units beyond the starting of the LES simulation. The normalized spectrum $E_u'(k)$ computed using DNS and LES data are quite close to each other.~\citet{Kumar:PRE2014} and  \citet{Verma:NJP2017} had shown earlier that turbulent RBC exhibits Kolmogorov's 5/3 scaling. Here we show that the LES too exhibits this scaling. The kinetic energy flux $\Pi_u(k)$  is constant in the inertial range ($k=[10,70]$), consistent with the constancy of $E_u'(k)$ observed for those wavenumbers. 

In Fig.~\ref{fig:E_th(k)_pi_1e8} we plot the entropy spectrum and flux  using the LES and DNS data at $t=30$. Note that LES, similar to DNS, captures the bi-spectrum of $E_{\theta}(k)$ quite well. Here, the upper branch exhibits $k^{-2}$ spectrum, whereas the lower branch is fluctuating. Mishra and Verma~\cite{Mishra:PRE2010} and Pandey et al.~\cite{Pandey:PRE2014} had shown that the upper $k^{-2}$ branch is constituted by dominant temperature modes $\theta(0,0,2n)$, which are approximately $-1/{2\pi}{n}$. Furthermore, the temperature modes in these two branches interact in such a way so as to yield constant entropy flux in the inertial regime. As shown in Fig.~\ref{fig:E_th(k)_pi_1e8}(b), $\Pi_{\theta}(k)$ obtained through LES exhibitsa  similar behaviour as DNS. Thus, the role of  dominant temperature modes and the interactions among these modes is captured quite well by our LES.

The above results describe similarities between the LES and DNS results for the global and spectral quantities.  We find that LES also captures the real space profile of DNS quite well. This is evident in Figure~\ref{fig:iso_1e8}(a,b) that exhibit the isosurfaces of temperature obtained in DNS and LES at time $t=27$ units. Note the similarity between the resolved structures in the two figures. Thus, the evolution of resolved scales in LES is quite similar to that in DNS.

These detailed comparisons between DNS and LES show that the present LES scheme is quite robust for simulating turbulent thermal convection.

\section{LES of RBC at very high Rayleigh numbers}
\label{sec:very_high_Ra}

\begin{figure}
\begin{center}
\includegraphics[scale=0.35]{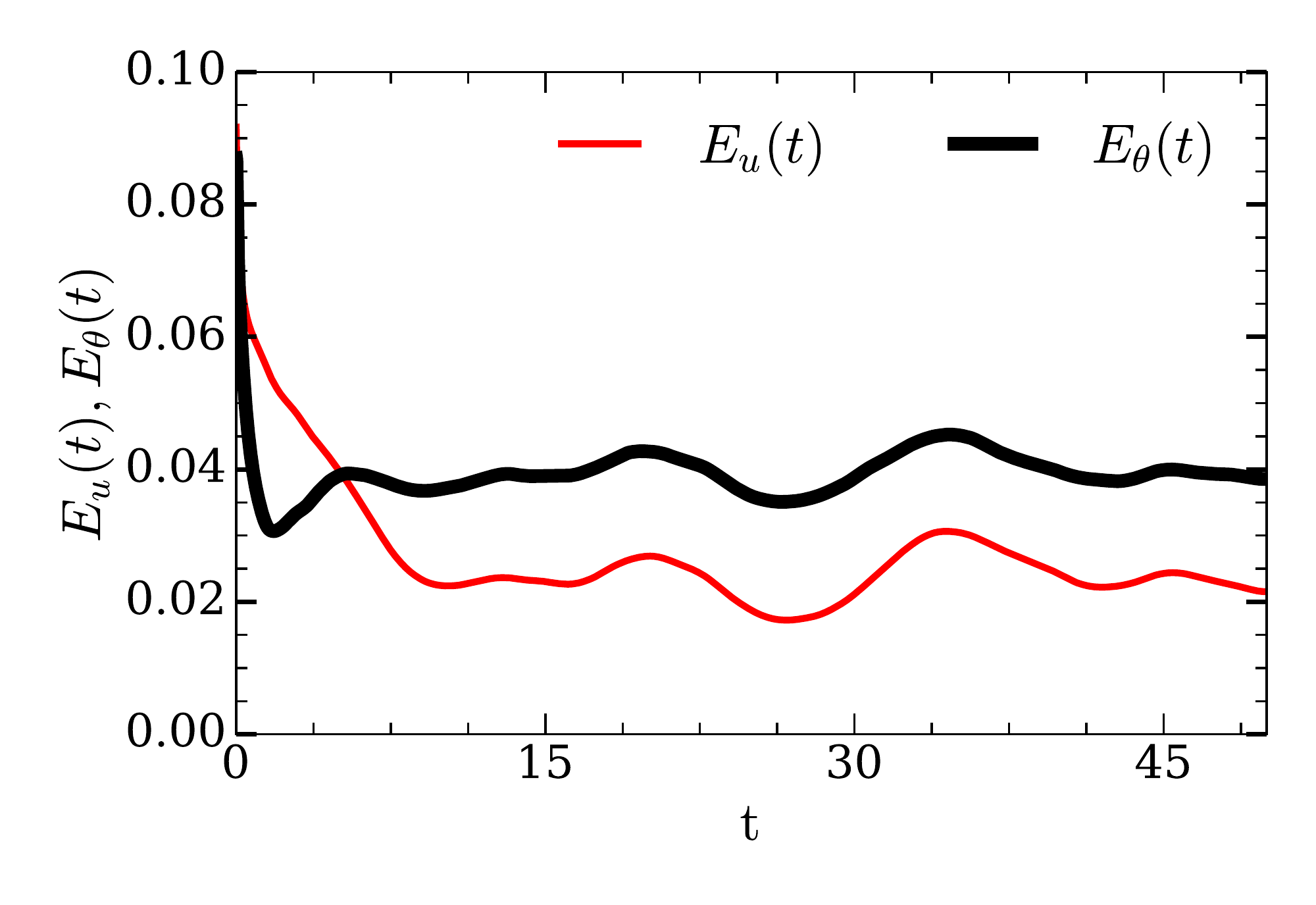}
\end{center}
\caption{ For a LES of RBC with $\mathrm{Ra=10^{18}}$: temporal evolution of total kinetic energy $E_u(t)$ and entropy $E_\theta(t)$..}
\label{fig:EhRa} 
\end{figure}

Buoyed by the success of the present LES scheme, we extended the LES runs to larger Rayleigh numbers. Here we present some of the results at $\mathrm{Ra} = 10^{18}$. We observe that our runs converge quite well.  Note that DNS of extreme  Ra is very difficult; the maximum Ra achieved in a 3D DNS is approximately $10^{12}$~\cite{Stevens:JFM2011, Verma:NJP2017}.  In the following we describe salient results obtained using our LES for $\mathrm{Ra} = 10^{18}$.

\begin{figure}
\begin{center}
\includegraphics[scale=0.35]{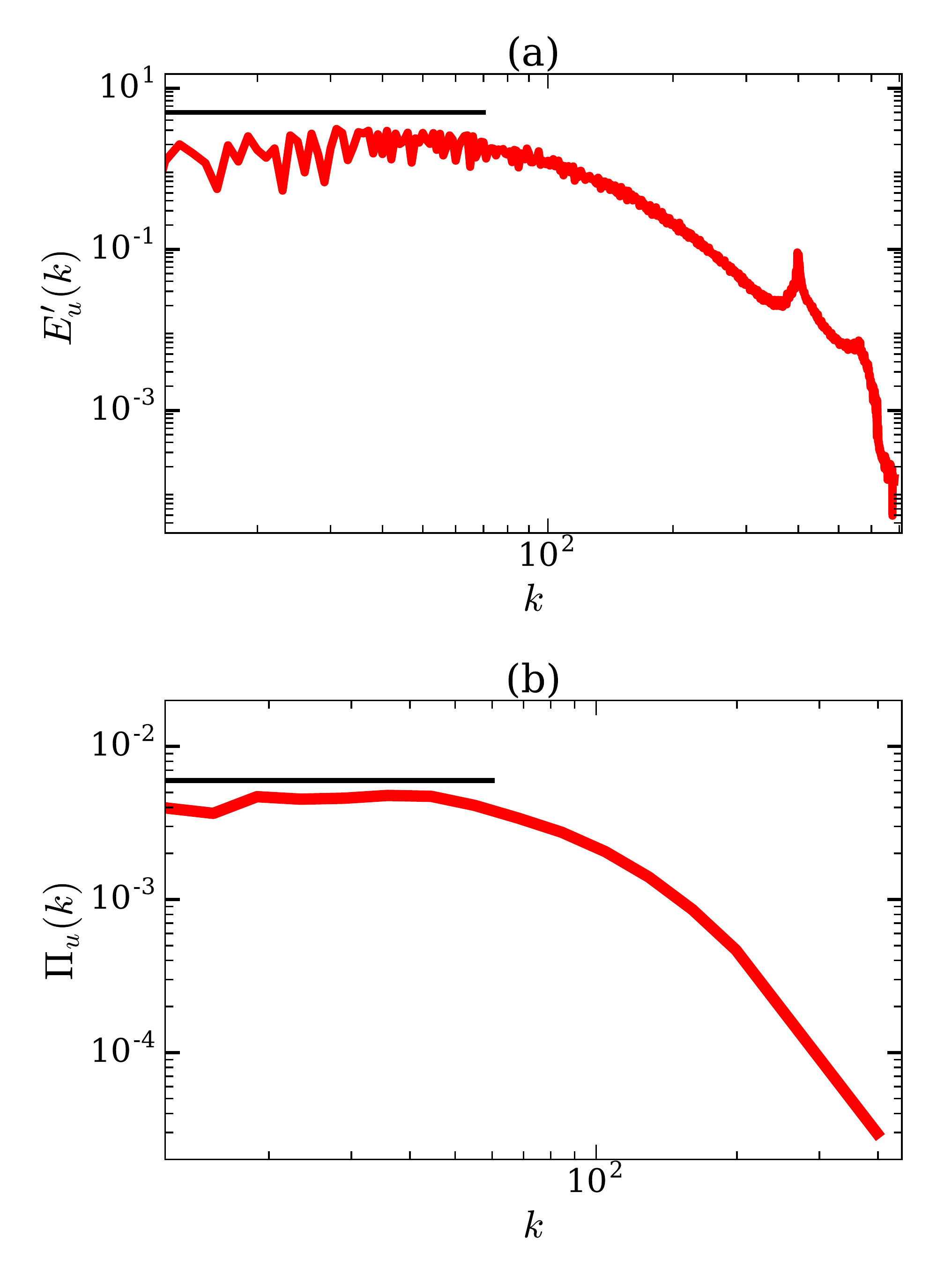}
\end{center}
\caption{(a) Kinetic energy spectrum and (b) Kinetic energy Flux for LES at $\mathrm{Ra=10^{18}}$.}
\label{fig:E_s-pihRa} 
\end{figure}

Starting from random initial conditions at $t=0$, we perform  LES on a $128^3$ grid and carry out our simulation till the system reaches a statistically stationary state.  In Fig.~\ref{fig:EhRa}, we plot the temporal evolution of total energy and entropy for  $\mathrm{Ra} = 10^{18}$. As shown in the figure, our LES converges quite nicely for such a large $\mathrm{Ra}$, which is quite surprising.

\begin{figure}
\begin{center}
\includegraphics[scale=0.35]{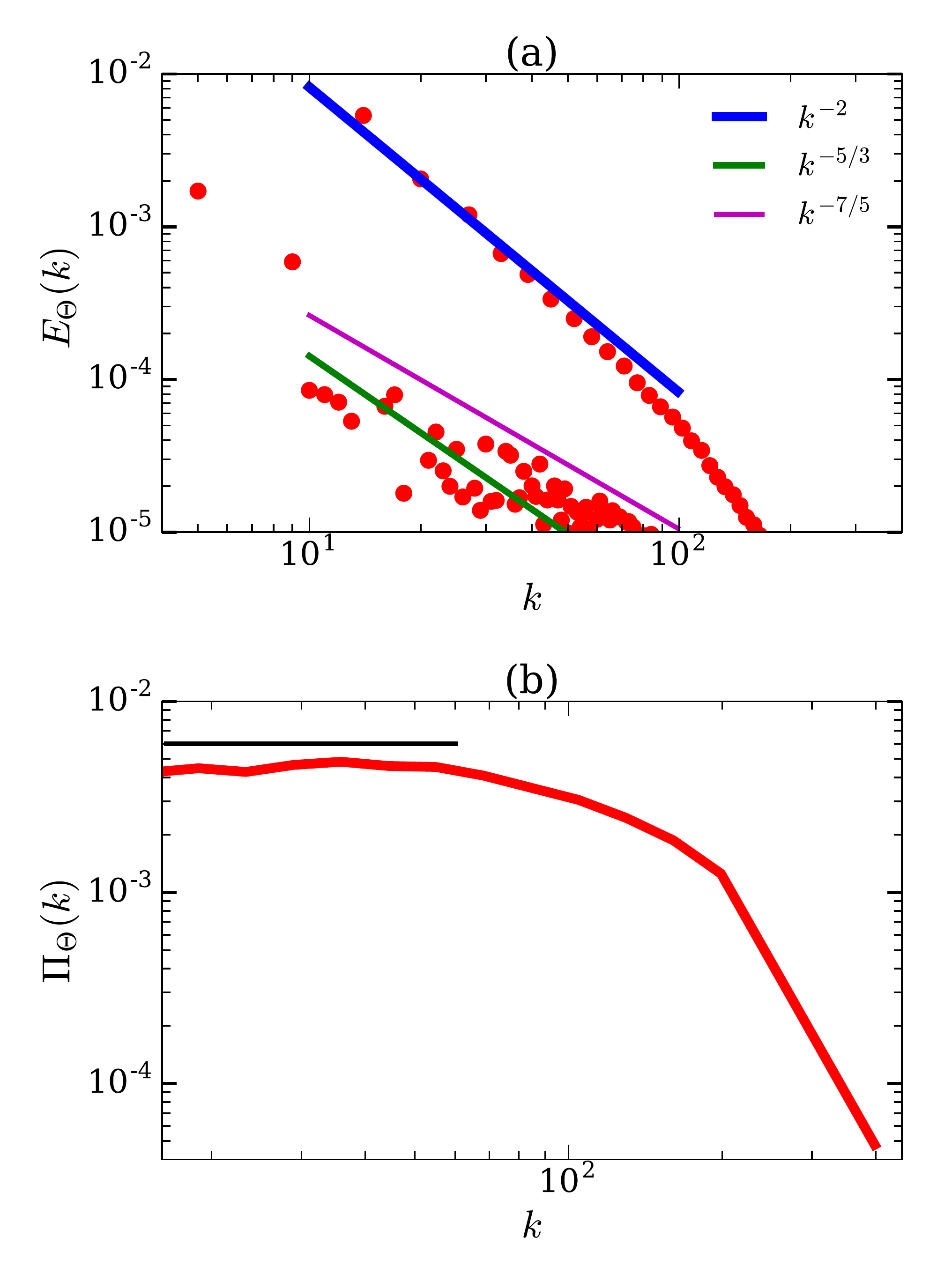}
\end{center}
\caption{(a) Entropy spectrum and (b) Entropy Flux for LES at $\mathrm{Ra=10^{18}}$.}
\label{fig:E_th-pihRa} 
\end{figure}

For this LES, we compute the spectra and fluxes of kinetic energy and entropy in the steady state. We plot these quantities in Figs.~\ref{fig:E_s-pihRa} and~\ref{fig:E_th-pihRa} for  $t=30$.   As shown in Fig.~\ref{fig:E_s-pihRa}(a,b), we obtain Kolmogorov's 5/3 scaling  for a narrow band of wavenumbers.  The kinetic energy flux is constant, as expected. 

 For the entropy spectrum, we observe a bi-spectrum as shown in Fig.~\ref{fig:E_th-pihRa}(a). The entropy flux (Fig.~\ref{fig:E_th-pihRa}(b)) is constant in the inertial range.   These results are similar to those obtained for $\mathrm{Ra}=10^8$ in the preceding section. Thus, the LES predicts  similar behaviour $\mathrm{Ra=10^8}$ and $\mathrm{Ra=10^{18}}$. 

\begin{figure}
\begin{center}
\includegraphics[scale=0.35]{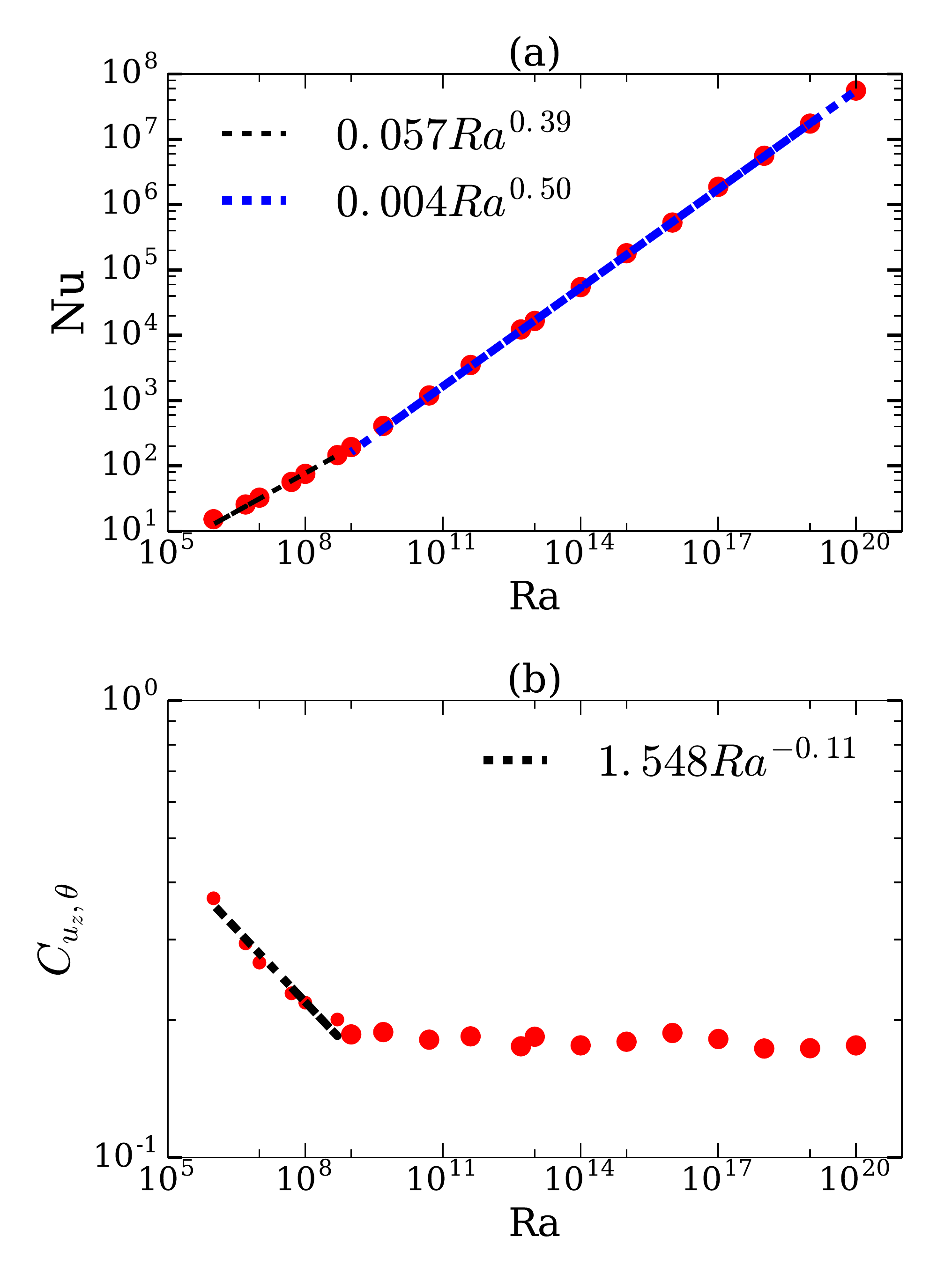}
\end{center}
\caption{(a) Scaling of Nusselt number $\mathrm{Nu}$ with Rayleigh number $\mathrm{Ra}$ and (b) Normalized velocity-temperature correlation. Both of these have been obtained using LES. Note the transition in the $\mathrm{Nu-Ra}$ scaling and an advent of the flattening of the correlation plot near $\mathrm{Ra=10^9}$.  }
\label{fig:Nu-Ra} 
\end{figure}

Now we report the   Nusselt number scaling  obtained using our LES data. Fig.~\ref{fig:Nu-Ra}(a) exhibits the Nu scaling on a log-log plot.  We observe that for $\mathrm{Ra} < 10^9$, $\mathrm{Nu\sim Ra^{\gamma_{1}}}$ with $\gamma_{1}=0.39\pm 0.01$, after which it makes a transition to $\mathrm{Nu\sim Ra^{\gamma_2}}$ with $\gamma_2=0.503 \pm 0.001$. Note however that earlier DNS and experiment results reveal that $\mathrm{Nu\sim Ra^{0.3}}$ for moderate Ra, and the exponent appears to increase~\cite{Niemela:Nature2000, Chavanne:PF2001, Lohse:ARFM2010} at very high Ra.  For example, \citet{He:PRL2012} argue that $\mathrm{Nu\sim Ra^{0.38}}$ beyond $\mathrm{Ra} \approx 5\times10^{14}$.   

As shown in Fig.~\ref{fig:Nu-Ra}, in our LES, a transition occurs  near $\mathrm{Ra} \sim 10^9$, beyond which we obtain $\mathrm{Nu\sim Ra^{1/2}}$, which is called the ultimate regime.  The  exponent of 1/2 was first predicted by \citet{Kraichnan:PF1962Convection} for very large Ra wherein  the boundary layer effects could be ignored.  Clearly our LES results for Nu-Ra scaling are inconsistent with the earlier DNS and experimental results. This discrepancy may be because of the inability of our SGS model to account for the boundary layers properly. Note that the boundary layer is a crucial component for the transition to the ultimate regime.  Nevertheless, it is heartening to see that our scheme is able to simulate RBC at huge Ra seamlessly.  A careful investigation of  the non-standard (and incorrect) Nu-Ra scaling obtained by our LES can provide insights into the flow properties of ultimate regime.

In addition to the Nusselt number, we also study the following normalized correlation function between vertical velocity and temperature fluctuations:
\be
C_{u_z, \theta}=\frac{{\big\langle u'_{z}\theta'\big\rangle}_V}{{\big\langle {u'_{z}}^{2}\big\rangle}_{V} {\big\langle \theta'^{2}\big\rangle}_{V}}.
\ee
\citet{Pandey:PF2016, Verma:NJP2017}, and \citet{Verma:PRE2012} have argued that the deviation of the Nu-Ra exponent from 1/2  to $\approx 0.30$ is due to nontrivial scaling of $C_{u_z, \theta}$ and $\theta$ fluctuations. It is argued  that $C_{u_z, \theta}\sim \mathrm{Ra^{-{0.22}}}$ for moderate $ \mathrm{Ra}$, and then it flattens out at very high Ra. We computed the above correlation function using LES data, and plot it as a function of Ra in Fig.~\ref{fig:Nu-Ra}(b). We observe that for lower Ra, $C_{u_z, \theta}\sim \mathrm{Ra^{-0.11}}$, consistent with $\mathrm{Nu} \sim \mathrm{Ra}^{1/2-0.11} \sim \mathrm{Ra}^{0.39}$.  But for large Ra, $C_{u_z,\theta} \sim \mathrm{const}$ (flat) that leads to $\mathrm{Nu} \sim \mathrm{Ra}^{1/2}$, consistent with the predictions of \citet{Kraichnan:PF1962Convection} for very large Ra.  These numerical observations are  consistent with the arguments made by \citet{Verma:NJP2017, Verma:PRE2012}. Thus, LES picks up the transition in $C_{u_z,\theta}$ quite well, albeit at lower Ra than expected. 

In summary, the present LES of turbulent convection has mixed success for very large Ra. It captures the  spectra and fluxes of the kinetic energy and entropy  quite well. However, the Nu-Ra scaling predicted by LES is inconsistent with the earlier experiments and direct numerical simulations. An encouraging point, however, is that we are able to reach very high Ra with the LES, and that it also captures the transition to the ultimate regime.
 
\section{Conclusions}
\label{sec:Conclusion}

 In this paper we present a SGS model for LES of turbulent thermal convection that  employs renormalized viscosity and thermal diffusivity.  Using this LES scheme we performed RBC simulations with free-slip and conducting plates for $\mathrm{Pr}=1$ and $\mathrm{Ra}=10^8$.   When we compare the LES results with those of DNS, we observe that the LES captures the evolution of total energy and entropy quite well. The spectra and fluxes of the kinetic energy and entropy and the isosurfaces of the temperature obtained through DNS and LES match with each other quite well. In addition, the LES is able to simulate RBC for Ra as large as $10^{18}$, and it also predicts a transition to the ultimate regime.  
 
 We remark here that the present LES scheme has a good scope of improvement to accommodate more generality. A realistic LES of thermal convection must capture the viscous and thermal boundary layers; this feature requires more sophisticated modelling of the viscosity and thermal diffusivity in the bulk and in the boundary layer.  Note that the local energy flux is expected to be different at different locations, specially in the bulk and in the boundary layer.  Hence, we need to model the energy flux $\Pi_u$ of Eq.~(\ref{eq:nu_total_RG}) locally. This can be computed using the third-order structure function~\citep{Frisch:book}. We plan to attempt such generalizations in near future.

\section*{ Acknowledgments} 
We thank  Fahad Anwer, Abhishek Kumar, Anando Chatterjee, Shashwat Bhattacharya, Manohar Sharma, and Mohammad Anas for useful discussions. The simulations were performed on the HPC system and Chaos cluster of IIT Kanpur, India. This work was supported by a research grant PLANEX/PHY/2015239 from Indian Space Research Organisation (ISRO), India. 

\bibliography{main}
\end{document}